# Valley-selective exciton bistability in a suspended monolayer semiconductor

Hongchao Xie[1], Shengwei Jiang[1,2], Jie Shan[1,2,3*], and Kin Fai Mak[1,2,3*]

[1] Laboratory of Atomic and Solid State Physics, Cornell University, Ithaca, New York 14853, USA

[2] School of Applied and Engineering Physics, Cornell University, Ithaca, New York 14853, USA

[3] Kavli Institute at Cornell for Nanoscale Science, Ithaca, New York 14853, USA

*E-mails: jie.shan@cornell.edu; kinfai.mak@cornell.edu

## Abstract

We demonstrate robust power- and wavelength-dependent optical bistability in fully suspended monolayers of $WSe_2$ near the exciton resonance. Bistability has been achieved under continuous-wave optical excitation at an intensity level of $10^3$ W/cm$^2$. The observed bistability is originated from a photo-thermal mechanism, which provides both optical nonlinearity and passive feedback, two essential elements for optical bistability. Under a finite magnetic field, the exciton bistability becomes helicity dependent, which enables repeatable switching of light purely by its polarization.

The extremely strong light-matter interaction and the presence of a new valley degree of freedom in monolayer transition metal dichalcogenides (TMD) semiconductors have attracted much interest in both fundamental science and new photonic and optoelectronic applications [1,2]. Monolayer TMD semiconductors are direct band gap materials [3,4]. Excitons, optically excited bound electron-hole pairs, are tightly bound in monolayer TMDs due to the weak dielectric screening in two dimensions [5-11], giving rise to extremely strong light-matter interaction near the exciton resonance [1]. Indeed, a single layer of $MoSe_2$ encapsulated by hexagonal boron nitride (hBN) has shown to reflect over 80% of incident light at the excitonic resonance [12,13]. The material has also exhibited optical bistability at high optical pump intensity of $10^6$ W/cm$^2$ [13]. On the other hand, monolayer TMD semiconductors possess two direct energy gaps located at the K and K' valleys of the Brillouin zone [1,2], which exclusively couple to light of opposite helicity [14-17]. Combining these two properties can, in principle, achieve the valley-selective optical nonlinearity or even bistability in monolayer TMDs. Optical bistability, the phenomenon of two well-discriminated stable states depending upon the history of the optical input, is the key concept for all-optical transistors, switches, logical gates, and memory [18]. Valley-selective exciton bistability will enable control of light not only by light intensity, but also by its polarization.

In this study, by fabricating fully suspended monolayer $WSe_2$, we demonstrate robust exciton bistability by continuous-wave (cw) optical excitation at a much lower intensity level of $10^3$ W/cm$^2$ compared to an earlier study [13]. The observed bistability is originated from a photo-thermal mechanism, which can provide both optical nonlinearity and internal passive feedback in a simple cavity-less structure. This is supported by detailed excitation wavelength and power dependence studies of the sample reflectance, as well as by numerical simulation including the temperature dependent optical response of monolayer $WSe_2$. Furthermore, under a finite magnetic field, the bistability becomes valley dependent, and controllable not only by light intensity but also by light helicity due to the exciton valley Zeeman effect [19-24]. Our results open up an exciting opportunity in controlling light with light using monolayer materials.

Suspended monolayer $WSe_2$ was fabricated by transferring monolayer $WSe_2$ flakes onto trenched $SiO_2$/Si substrates. Figure 1a shows the optical image and schematic side view of a representative monolayer sample suspended over a circular trench of 8 μm in diameter and 600 nm in depth. Figure 1b is its reflectance spectrum at temperature ranging from 4 K to 330 K. The reflectance $R/R_0$ at any given wavelength was determined as the ratio of the reflected power from the suspended sample $R$ to the reflected power from the bare trench $R_0$ under normal incidence. It exhibits a sharp exciton resonance with over 80% reflectance contrast at low temperatures. The spectra can be simulated by considering light propagation in the $WSe_2$/vacuum/Si multilayer structure with the $WSe_2$ optical response described by a single Lorenzian arisen from the exciton resonance (Fig. 1c). The small discrepancy around 720 nm at 4 K is due to the contribution of charged exciton to the optical response of monolayer $WSe_2$ [25-27], which has been neglected in the simulation for simplicity. The exciton peak wavelength and linewidth employed in the simulation are shown in Fig. 1d as a function of temperature. In the low temperature limit, the exciton is peaked at 709 nm (1.75 eV) with a linewidth of ~ 7 meV. With increasing temperature, the exciton resonance redshifts and broadens. This behavior is well described by the exciton-phonon interaction (solid lines) with an average phonon energy of ~ 16 meV and LO-phonon energy of ~ 30 meV [28, 29]. These values are in good agreement with the values reported for WSe$_2$ [28]. (See Methods for more details on the sample preparation and analysis of the reflectance spectra.)

We study the optical nonlinearity in monolayer WSe$_2$ near the exciton resonance at 4 K. Figure 2 shows the power dependence of the reflectance for cw excitation tuned through the exciton resonance in sample #1. The beam size on the sample is typically about 1 μm. The results are similar up to ~ 100 K, for which thermal broadening is insignificant. They are also reproducible in all samples studied here except small variations in the exciton resonance wavelengths and linewidths. Results at selected wavelengths, marked as vertical dashed lines on the experimental reflectance spectrum (Fig. 2a) and on the corresponding simulated absorbance spectrum (Fig. 2b), are shown in Fig. 2c. At 708 nm (blue-detuned from the exciton resonance), the reflectance increases with increasing power. Onset of nonlinearity occurs at a power level as low as 1 μW. At all other wavelengths (red-detuned from the exciton resonance), the reflectance exhibits jumps at critical powers and a hysteretic behavior with power sweeps. These are signatures of optical bistability. Moreover, both the critical powers and the size of the bistability region increase for excitation at longer wavelengths. We note that several μW is sufficient to generate optical bistability in suspended monolayers here, in contrast to ~ 10 mW in monolayer TMDs encapsulated in hBN under similar experimental conditions [13].

To understand the origin of the observed optical bistability, we monitor the reflectance spectrum with a low-power super-continuum probe of suspended samples that are pumped by cw excitation at a fixed power of 100 μW. Figure 3a, b are the contour plots of the reflectance as a function of the pump (left axis) and probe (bottom axis) wavelengths of sample #2. Figure 3a corresponds to a forward sweep of the pump wavelength (i.e. increasing wavelength), and Fig. 3b, a backward sweep (i.e. decreasing wavelength). The bright features along the diagonal dashed lines are the residual of the pump intensity. For the forward sweep, a clear exciton resonance is initially observed at 715 nm with a width of ~ 10 meV. As the pump wavelength approaches the exciton resonance, a significant redshift of the exciton by over 20 nm accompanied with a broadening by ~ 100 meV is observed. A sudden "recovery" of the exciton resonance follows when the pump wavelength increases beyond a critical value of ~ 745 nm. In contrast, for the backward sweep of the pump wavelength, the exciton resonance suddenly redshifts to ~ 738 nm at a lower critical pump wavelength of ~ 735 nm. The exciton gradually returns to the original position as the pump wavelength further decreases. We summarize the pump wavelength dependence of the exciton peak and linewidth in Fig. 3c and 3d, respectively. These values were extracted from the analysis of the reflectance spectra (Methods). Clear hysteresis in both the peak wavelength and the linewidth can be observed. Figure 3a, b also show that reflectance at probe wavelengths off the exciton resonance (or far above the resonance at 450 nm, not shown) remains largely independent of pump. Since the reflectance from the multilayer structure is very sensitive to the vacuum cavity length due to optical interference, our result indicates a negligible pump-induced vertical movement of the suspended membrane.

The significant pump-induced exciton redshift and broadening near the bistability region suggests the importance of a photo-thermal mechanism. We estimate from Fig. 1d that a temperature rise of ~ 300 K is required to explain the observed exciton resonance shift under near-resonance (740 nm) excitation at 100 μW. Such a large temperature rise, as we discuss below, is plausible due to the low thermal conductance of suspended samples. As the sample temperature rises under optical pumping due to the photo-thermal effect, the exciton resonance "runs" away from the pump wavelength for a blue-detuned pump, which in turn decreases the sample absorbance at the pump wavelength and reduces the temperature rise. No cumulative effect can build up. In contrast, the exciton resonance "runs" towards the pump wavelength for a red-detuned pump, which increases the sample absorbance at the pump wavelength and, in turn, causes a further increase in the sample temperature and more redshift towards the pump wavelength. This process is cumulative, i.e. a positive feedback. This internal passive feedback, together with optical nonlinearity, leads to optical bistability in monolayer $WSe_2$ (Fig. 4a). For a red-detuned pump under a forward power sweep, the system becomes unstable and switches suddenly from a low-temperature state to a high-temperature state when the pump power $P$ exceeds a critical value $P_f$. Under a backward power sweep, the system starts from the high-temperature state, and can remain there even below $P_f$. Only when the pump power decreases below a lower critical value $P_b$, optical pumping can no longer keep the system in the high-temperature state, and it switches suddenly to the low-temperature state. A hysteretic bistable behavior occurs.

We now turn to a more quantitative discussion of optical bistability from the photo-thermal effect. We can express the average sample temperature under optical excitation at wavelength $\lambda$ and power $P$ as

$$T(P,\lambda) = T_0 + \frac{A(T,\lambda)L(T,\lambda)P}{4\pi\kappa(T)}, \tag{1}$$

where $T_0$, $A(T,\lambda)$ and $\kappa(T)$ are, respectively, the initial temperature, optical absorbance and sheet thermal conductance of the sample; $L(T,\lambda)$ is the local field factor. Optical bistability requires the existence of more than one solution to $T$ for a given $P$ (Fig. 4b). Therefore in addition to normal regions of monotonically increasing $T$ with $P$, there must exist a region of decreasing $T$ with $P$, i.e. $\frac{dT}{dP} \leq 0$. Ignoring the weak temperature dependence of the local field factor (which is mainly determined by the substrate properties) and the sample thermal conductance for simplicity, we obtain from Eqn. (1) a criterion for optical bistability

$$\frac{dA}{dT} \gtrsim \frac{A}{T-T_0}. \tag{2}$$

Since $\frac{dA}{dT}$ is negative (positive) on the blue (red) side of the exciton resonance (Fig. 1b, c) and $\frac{A}{T-T_0}$ is always positive, optical bistability can only occur for sufficiently red-detuned optical pumping. This is fully consistent with our experimental observation and the above intuitive picture for the phenomenon.

Next we compare in details the predictions of the photo-thermal effect and the experimental observations. To this end, we solve Eqn. (1) numerically for $T(P,\lambda)$. For simplicity we have used a constant thermal conductance ($\kappa \approx 5\times10^{-9}$ W/K). Figure 4b shows $T(P)$ for two representative wavelengths. At 708 nm (blue-detuned from the exciton resonance), $T$ increases monotonically with $P$. In contrast, at 728 nm (red-detuned from the exciton resonance), temperature instability occurs at critical power $P_f$ and $P_b$ for forward and backward power sweeps, respectively. The system is bistable between these two powers. Figure 4c is $T(\lambda)$ for three representative powers (1, 10 and 100 μW). Temperature instability is observed only at 10 and 100 μW. Again, the system is bistable between the critical wavelength $\lambda_f$ and $\lambda_b$, at which temperature instability occurs for forward and backward wavelength sweeps, respectively. Moreover, the bistability region shifts towards more red-detuned wavelengths for larger powers. Now equipped with the sample temperature, we can simulate its reflectance using the measured temperature dependent absorbance. Figure 2d is the simulated power dependence of the sample reflectance. The simulation reproduces the overall trend of the experiment (Fig. 2c) for all wavelengths with critical powers within a factor of 2 − 3 of the measured values. Figure 5d - 5f are the simulated wavelength dependence of the sample reflectance at a fixed power of 1, 10 and 100 μW, respectively. The agreement between simulation and experiment (Fig. 5a - 5c) is also very good including the critical wavelengths. We have also included in Fig. 3e,f simulation for the pump-probe experiment (Fig. 3c,d). Again, reasonably good agreement is seen for the exciton characteristics as a function of pump wavelength. We note that there is only one free parameter (the sample thermal conductance) in the simulation. The value of $5\times10^{-9}$ W/K was chosen to fit the entire set of the experimental observations. It is consistent with the measured [30, 31] and calculated [32] thermal conductivity of WSe$_2$. The low thermal conductance of suspended WSe$_2$ here is primarily responsible for the low critical powers required for optical bistability. We conclude that the photo-thermal effect captures the main features of optical bistability in suspended monolayer WSe$_2$ despite the crude assumptions made in the simulation, including the temperature-independent thermal conductance and spatially uniform heating.

Finally, we demonstrate the unique valley-selective optical bistability and control of light by its helicity in suspended monolayer WSe$_2$ (sample #3). The idea relies on the valley-dependent optical response of monolayer TMDs [1, 2, 14-17, 19-24, 26], in which the two degenerate direct gaps at the K and K' valleys of the Brillouin zone couple exclusively to the left ($\sigma^+$) and right ($\sigma^-$) circularly polarized light, respectively. Since the K and K' valleys are time-reversal copies of each other, the valley degeneracy can be lifted by applying an out-of-plane magnetic field (i.e. the exciton valley Zeeman effect [19-24]). This is shown in Fig. 6a for the helicity-resolved reflectance spectrum measured under a magnetic field of 8 T. The exciton resonance is split by ~ 2 meV, in agreement with the reported magnitude of the exciton valley Zeeman effect [19-24]. We study in Fig. 6b the optical nonlinearity for the $\sigma^+$ and $\sigma^-$ excitation at a red-detuned wavelength (710 nm, marked by the dashed vertical line in Fig. 6a). Optical bistability is observed for both circular polarizations with larger critical powers and a wider hysteresis loop for the $\sigma^+$ excitation. This is consistent with the fact that 710 nm corresponds to a larger redshift from the exciton resonance for the $\sigma^+$ excitation. Moreover, there exists a finite range of power between the two bistability regions (marked by two dotted vertical lines). This can be utilized for repeatable switching of the sample reflectance by light helicity, as illustrated in the power dependence of the magnetic circular dichroism (MCD) (Fig. 6c).

Now consider the forward power sweep. At a power inside the bistability region of $\sigma^+$, switching from $\sigma^+$ to $\sigma^-$ switches the system from the high- to the low-reflectance state. But subsequent switching of the light helicity cannot bring it back to the high-reflectance state since the (high-temperature) low-reflectance state is the low-energy state of the system. The measured MCD, averaged over many modulation cycles, thus vanishes at this power. Similar argument can be made for the backward power sweep. This is consistent with the result of Fig. 6c if we take into account the broadened hysteresis loops. We thus choose a power between the bistability regions (39 µW) at 710 nm to demonstrate repeatable light switching by pure helicity in Fig. 6d. In the upper panel, the light helicity is modulated by a liquid crystal modulator at 1 Hz. Repeatable switching of the sample reflectance by ~ 30 % is achieved (lower panel). (Note the delayed response in the reflectance when switching from $\sigma^+$ to $\sigma^-$ is due to the slow response of our liquid crystal modulator rather than limited by the intrinsic response of the material.) Although the unique valley-dependent optical bistability and helicity-controlled optical switching can only be achieved under a finite magnetic field here, it is, in principle, also feasible to achieve such a phenomenon under zero magnetic field utilizing the magnetic proximity effect. Recent experimental studies have shown strong magnetic proximity coupling of monolayer TMDs to magnetic insulators with a proximity exchange field exceeding 10 T [33, 34]! Our study has thus opened up exciting opportunities in controlling light with light, including wavelength, power, and helicity, using monolayer materials.

**Methods**

**Device fabrication.** Monolayer WSe$_2$ samples were mechanically exfoliated from synthetic bulk crystals onto polydimethylsiloxane (PDMS). They were then transferred onto drumhead trenches on Si substrates with a 600-nm oxide layer. Drumhead trenches were etched using reactive ion etching (RIE). The sample thickness was determined by their reflectance contrast and photoluminescence spectrum. Five suspended samples over drumhead trenches of 8 µm in diameter and 600 nm in depth were studied.

**Optical measurements.** Suspended WSe$_2$ samples were measured in an optical cryostat. Light was coupled into and out of it using a high numerical aperture microscope objective under normal incidence. The beam size on the sample is typically about 1 µm. For reflectance spectrum measurements, broadband radiation from a super-continuum light source and a spectrometer equipped with a liquid nitrogen cooled charge coupled device (CCD) were used. The power on the sample was limited to < 1 µW to avoid any nonlinear effects. For the cw excitation measurement, a tunable cw Ti-sapphire laser and a Si avalanche photodiode were employed. The laser power on the sample was controlled by rotating a half waveplate sandwiched between two linear polarizers. It was limited to 250 µW to avoid sample damage. In the pump-probe measurement, the reflected pump beam was filtered by a crossed linear polarizer. In the helicity-resolved measurement, a liquid crystal phase plate was used to change the incident light polarization from linear to circular. For the MCD measurement and the reflectance switching experiment, the light helicity was modulated by electrically modulating the liquid crystal phase plate at several Hz's and the reflected light was detected by an avalanche photodiode. The MCD signal was recorded with a lock-in amplifier and the time trace of the reflected light was recorded with an oscilloscope.

**Analysis.** The sample reflectance $R/R_0$ at a given wavelength $\lambda$ or frequency $\omega = 2\pi c/\lambda$ ($c$ denoting the speed of light) is characterized by the ratio of the reflected power from the suspended sample $R$ to that from the bare trench $R_0$ under normal incidence. These two quantities can be expressed through the parameters of the WSe$_2$/vacuum/Si multilayer structure and the optical properties of its constitutes as [6]

$$R(\omega) = \left|-\frac{C-D[1-\tilde{A}(\omega)]}{C+D[1+\tilde{A}(\omega)]}\right|^2, \quad R_0(\omega) = \left|-\frac{C-D}{C+D}\right|^2. \tag{3}$$

Here we have assumed the incident power to be unit. The coefficients $C$ and $D$ are related to the refractive index $n_\zeta$, thickness $d_\zeta$, and phase shift $\delta_\zeta = n_\zeta \omega d_\zeta/c$ in the vacuum ($\zeta = v$) and Si ($\zeta = Si$) layers:

$$\binom{C}{D} = \begin{pmatrix} \sin\delta_v & in_{Si}\cos\delta_v \\ i\cos\delta_v & n_{Si}\sin\delta_v \end{pmatrix} \binom{\sin\delta_{Si} + in_{Si}\cos\delta_{Si}}{i\cos\delta_{Si} + n_{Si}\sin\delta_{Si}}. \tag{4}$$

The complex absorbance $\tilde{A}(\omega)$ of monolayer WSe$_2$ is modeled by a single Lorentzian arisen from the exciton resonance

$$\tilde{A}(\omega) = \frac{A_0}{1 - i2\hbar(\omega-\omega_0)/\gamma}. \tag{5}$$

Here $A_0$ is the area of the Lorentzian, $\omega_0$ is the resonance frequency, and $\gamma$ is the total exciton linewidth. The sample absorbance $A$ in Eqns. (1) and (2) is the real part of $\tilde{A}$, and the local field factor $L = \left|1-\frac{C-D[1-\tilde{A}]}{C+D[1+\tilde{A}]}\right|^2 \approx \left|1-\frac{C-D}{C+D}\right|^2$ is primarily determined by the parameters of the substrate and is assumed to be temperature independent. As shown in Fig. 1c, the reflectance spectrum at 4 K can be fitted well using $A_0 \approx 0.85$, $\hbar\omega_0 \approx 1.75$ eV and $\gamma \approx 7$ meV. The temperature-dependent exciton peak position $\omega_0(T)$ and linewidth $\gamma(T)$ (Fig. 1d) are well described by [28, 29]

$$\omega_0(T) = \omega_0(0) - S\langle\omega_D\rangle\left[\coth\left(\frac{\langle\hbar\omega_D\rangle}{2k_BT}\right) - 1\right], \tag{6}$$

$$\gamma(T) \approx \gamma(0) + \frac{\gamma_0}{Exp(\hbar\omega_D/k_BT)-1}. \tag{7}$$

Here $\hbar\omega_0(0) \approx 1.75$ eV is the exciton resonance at zero temperature, $\langle\hbar\omega_D\rangle \approx 16$ meV and $\hbar\omega_D \approx 30$ meV are the average phonon energy and LO-phonon energy of WSe$_2$, respectively, $S \approx 2.25$ is the coupling parameter, $k_B$ is the Boltzmann constant, $\gamma(0) \approx 7$ meV is the linewidth at zero temperature, and $\gamma_0 \approx 76$ meV is the characteristic linewidth corresponding to exciton-phonon scattering [28].

**Figures**

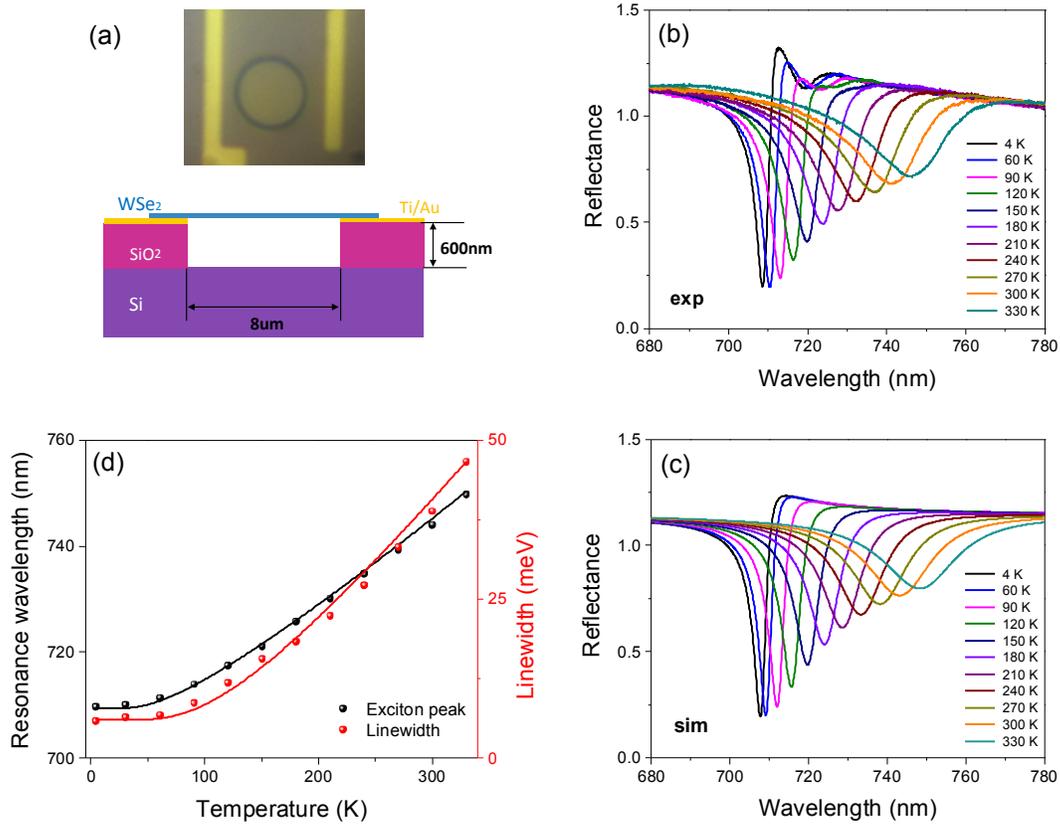

**Figure 1.** (a) Schematic side view of a suspended monolayer WSe$_2$ sample over a drumhead trench of 8 μm in diameter and 600 nm in depth on a Si substrate. The substrate has a 600-nm oxide layer and pre-patterned Ti/Au electrodes. The inset is an optical image of a sample illuminated by white light. Gold bars are electrodes and the dark ring is the edge of the trench. (b) Reflectance spectrum measured at 4 - 330 K. A sharp exciton resonance is observed at low temperatures. The resonance redshifts and broadens with increasing temperature. (c, d) Simulated reflectance spectrum at 4 - 330 K (c) with the extracted exciton peak wavelength and linewidth (symbols) as a function of temperature (d). Solid lines in (d) are fits to Eqn. (6) and (7).

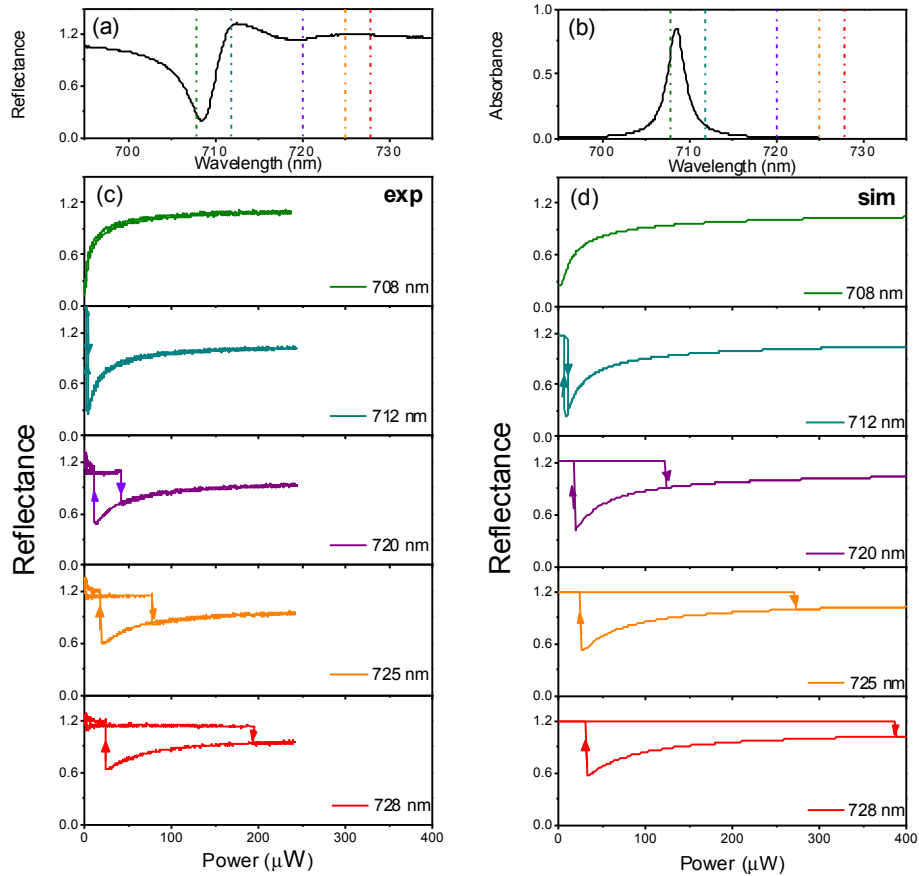

**Figure 2.** (a) Reflectance spectrum of sample #1 at 4 K. (b) Absorbance spectrum described by a single Lorentzian that best fits the reflectance spectrum of (a). (c, d) Reflectance at representative wavelengths (708, 712, 720, 725 and 728 nm) under forward and backward power sweeps from experiment (c) and simulation (d). The wavelengths are marked by vertical dashed lines in (a, b).

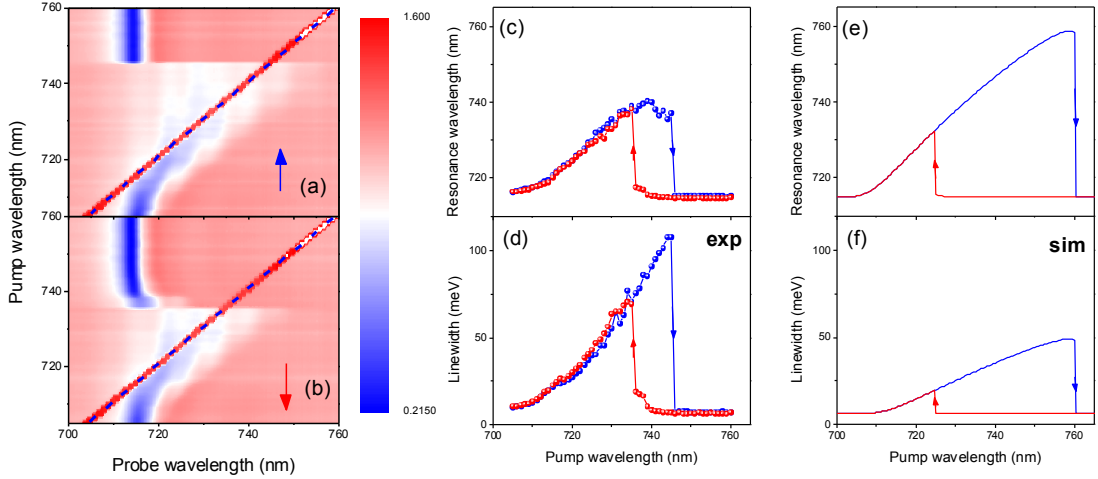

**Figure 3.** (a, b) Contour plots of reflectance of sample #2 at 4 K as a function of probe wavelength (bottom axis) and pump wavelength (left axis). The pump wavelength is swept forward (i.e. increases) in (a) and backward (i.e. decreases) in (b). The pump power is fixed at 100 μW. The diagonal dashed lines are the residual pumps. (c, d) The extracted exciton peak wavelength (c) and linewidth (d) from the experimental reflectance spectra. (e, f) Simulation of the parameters shown in (c, d) based on a photo-thermal effect.

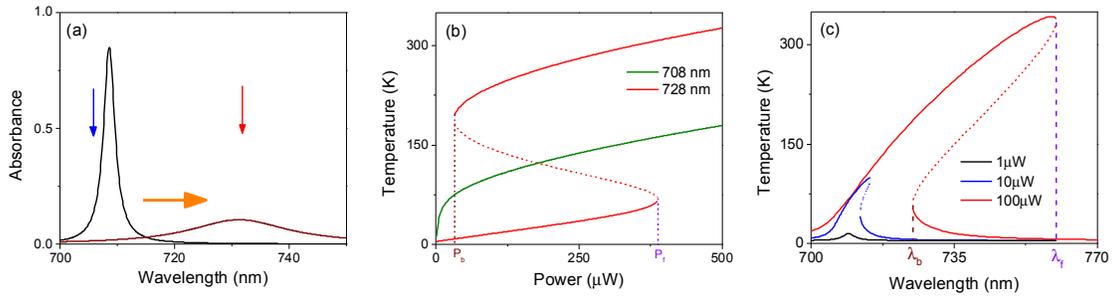

**Figure 4.** (a) Photo-thermal effect: the exciton resonance redshifts under optical excitation for both blue- and red-detuned optical excitation from the exciton resonance. For the red-detuned optical excitation, the exciton resonance runs towards the pump and the effect is cumulative. This internal passive feedback together with nonlinearity can lead to optical bistability. (b) Simulated average sample temperature as a function of excitation power at 708 nm (blue-detuned from the exciton resonance) and 728 nm (red-detuned from the exciton resonance). $P_f$ and $P_b$ are the critical power for temperature instability under forward and backward power sweeps, respectively. (c) Simulated average sample temperature as a function of wavelength at 1, 10 and 100 μW. $\lambda_f$ and $\lambda_b$ are the critical wavelengths for temperature instability under forward and backward wavelength sweeps, respectively. The region between the critical powers (wavelengths) is bistable.

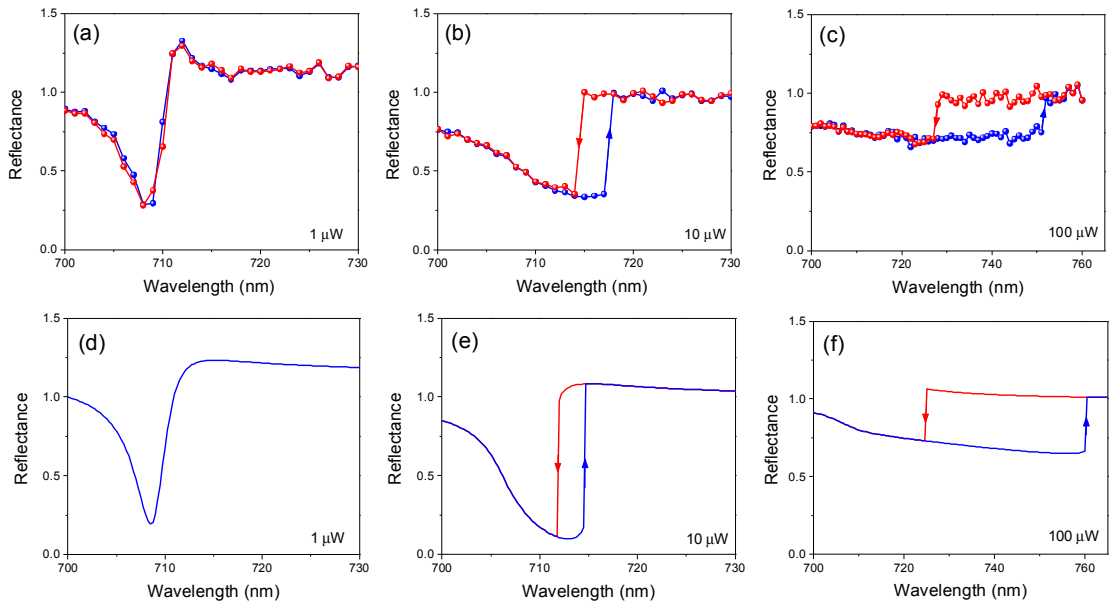

**Figure 5.** Wavelength dependence of the sample reflectance for forward (blue) and backward (red) wavelength sweeps. (a, b, c) are experiment and (d, e, f) are simulation. The excitation power is fixed at 1 µW (a, d), 10 µW (b, e) and 100 µW (c, f).

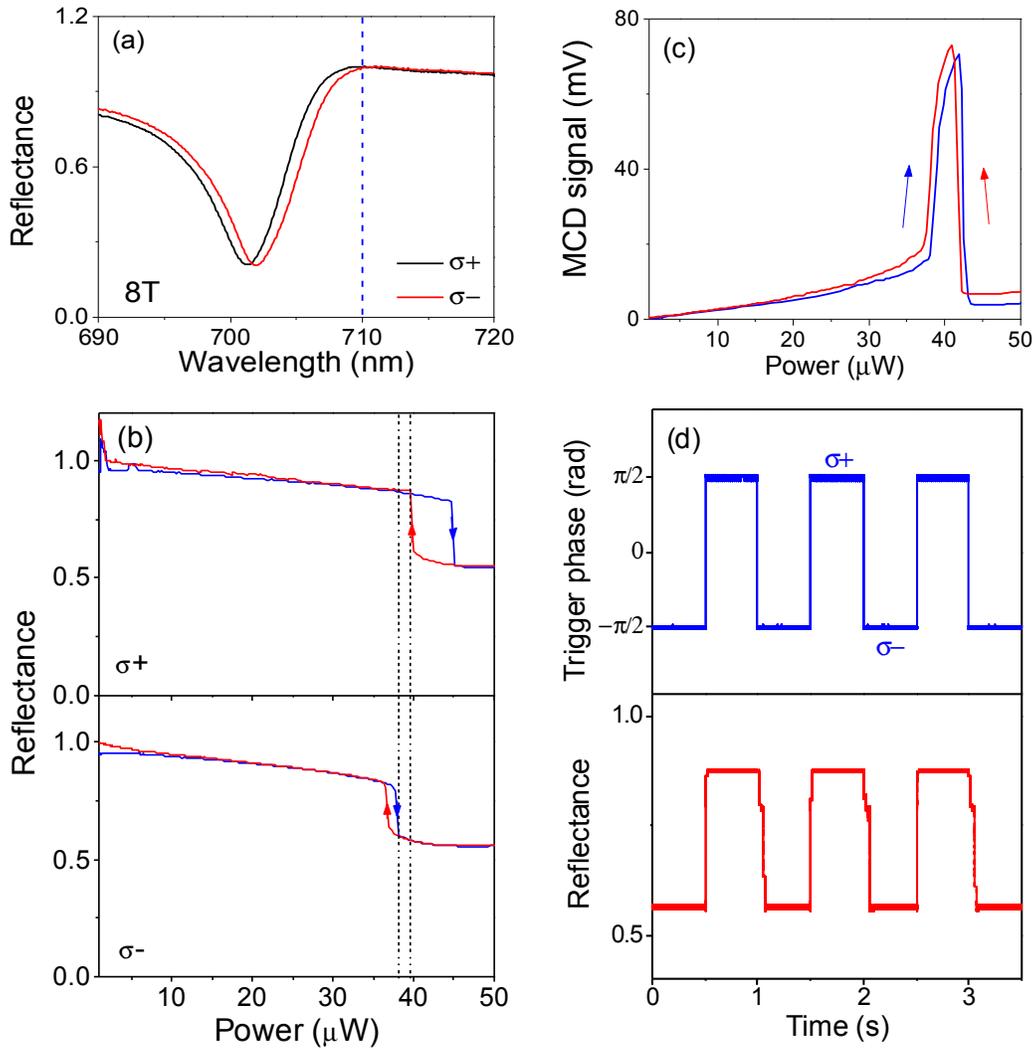

**Figure 6.** (a) Helicity-resolved reflectance of sample #3 at 4 K under an out-of-plane magnetic field of 8 T. The K and K' valley excitons are split by ~ 2 meV. The vertical dashed line corresponds to 710 nm. (b) Helicity-resolved reflectance under power sweeps at 710 nm. There exists a finite power range between two bistability regions marked by vertical dotted lines. (c) Magnetic circular dichroism (MCD) under forward and backward power sweeps at 710 nm. (d) Helicity-controlled optical 'switching' in real time at 710 nm and 39 μW. The helicity of light is modulated at 1 Hz by a liquid crystal modulator (upper panel) and the sample reflectance follows the helicity modulation with a change of reflectance by ~ 30% (lower panel).